\documentclass[aps,nofootinbib,showpacs,showkeys,preprintnumbers,amsmath,amssymb]{revtex4}		

\usepackage{amsmath,amsfonts,amssymb,amsthm,bm,bbm,bbold,braket,color,dsfont,fancybox,hyperref,
srcltx,subfigure,slashed,ucs,yfonts,cancel,xfrac,verbatim,xcolor}
\usepackage{floatrow}
\usepackage{graphicx}
\usepackage[outercaption]{sidecap} 
\usepackage[inline,shortlabels]{enumitem}
\usepackage{nicefrac}
\usepackage{multirow}
\DeclareMathAlphabet{\mathpzc}{OT1}{pzc}{m}{it}
\definecolor{gold}{rgb}{1,0.8,0}
\definecolor{nara}{rgb}{1,0.4,0.1}
\definecolor{goldo}{rgb}{1,0.7,0}
\definecolor{greeno}{rgb}{0,0.8,0}
\def\bes{\begin{subequations}}
\def\ees{\end{subequations}}
\def\be{\begin{equation}}
\def\ee{\end{equation}}
\def\bea{\begin{eqnarray}}
\def\eea{\end{eqnarray}}
\def\ba{\begin{eqnarray}}
\def\ea{\end{eqnarray}}
\def\bear{\begin{array}}
\def\eear{\end{array}}
\newcommand{\txty}[1]{\textrm{\tiny{#1}}}
\newcommand{\bpm}{\begin{pmatrix}}
\newcommand{\epm}{\end{pmatrix}}
\newcommand{\BM}{\left(\begin{array}}		
\newcommand{\BMC}{\left[\begin{array}}		

\newcommand{\EM}{\end{array}\right)}		
\newcommand{\EMC}{\end{array}\right]}		
\newcommand{\hs}[1]{\hspace{#1mm}}	
\newcommand{\SB}{\overline}
	% ...

\newcommand{\mM}{\mathcal{M}}

\newcommand{\mO}{\mathcal{O}}

\newcommand{\cL}{\mathcal{L}}	

\newcommand{\hc}{\textrm{h.c.}}

\newcommand{\com}[1]{}
\newcommand{\eV}{\,\textrm{eV}}

\newcommand{\GeV}{\,\textrm{GeV}}

\newcommand{\KeV}{\,\textrm{KeV}}

\newcommand{\m}{\left[\textrm{m}\right]}
 
\newcommand{\MeV}{\,\textrm{MeV}}

\newcommand{\TeV}{\,\textrm{TeV}}

\begin{document}

\title{Rare meson decays with three pairs\\ of quasi-degenerate heavy neutrinos.}
\author{Gast\'on Moreno $^{1,2}$}
\email{gaston.moreno@usm.cl}
\author{Jilberto Zamora-Saa $^{2,3}$}
\email{jzamorasaa@jinr.ru}

\affiliation{
$^{1}$ Department of Physics, Universidad T\'ecnica Federico Santa Mar\'ia, Valpara\'iso, Chile\\
$^{2}$ Centro Cient\'{\i}fico Tecnol\'ogico de Valpara\'{\i}so, Valpara\'iso, Chile\\
$^{3}$ Dzhelepov Laboratory of Nuclear Problems, Joint Institute for Nuclear Research, Dubna, Russia}

\begin{abstract}
We consider a scenario of Low Scale Seesaw where the masses of its heavy Majorana neutrinos are arranged in a pattern of three pairs of quasi degenerate ones, in the range of {$\mathcal{O}(1-6\GeV)$}. Since they can violate lepton number by two units, they contribute to rare decays of $D_s$ and $B_c$ mesons, providing, besides, conditions for maximal CP violation. We found that new phenomenology is possible depending on how many of on-shell pairs mediate these decays. In particular, we present new constraints on muon-heavy neutrinos mixing parameters.
\end{abstract}

\maketitle

\section{Introduction}
\label{intro}

Recent experiments have shown that neutrinos have non-zero  masses~\cite{Fukuda:1998mi,Eguchi:2002dm}, and that the mixing pattern between mass and flavour states is explained by the Maki-Nakagawa-Sakata Matrix, $U_\txty{MNS}$~\cite{Maki:1962mu}. Within the requisite that the matrix which diagonalize the whole neutrino mass matrix is unitary, current uncertainties in the elements of $U_\txty{MNS}$ allow a small range for mixing between SM flavours states and new sterile ones~\cite{Antusch:2008tz,Malinsky:2009gw,Dev:2009aw,Forero:2011pc,Antusch:2014woa}, which would imply a tiny interaction of the latter with SM particles. Likewise, as the $\theta_{13}$ angle of $U_\txty{MNS}$ is non zero~\cite{Beringer:1900zz,GonzalezGarcia:2012sz} it remain open the possibility that light neutrino sector could violate the CP symmetry; certainly, this is not enough to explain Baryogensis via Leptogenesis \cite{Strumia:2006qk}. Additionally, depending whether neutrino nature is Dirac or Majorana we would have three (Majorana) or one (Dirac) CP violation phases.~\cite{Bilenky:2004sn}. It has been proposed~\cite{Littenberg:1991ek,Pilaftsis:1997jf,Littenberg:2000fg,Dib:2000wm,Ali:2001gsa,Ivanov:2004ch,Helo:2010cw,
Cvetic:2012hd,Cvetic:2013eza,Cvetic:2014nla} that detection of rare decays of mesons of type $M \to \ell_1 \ell_2 M^\prime$ (with $M,M^\prime$ being mesons, whereas $\ell_{1},\ell_{2}$ are charged leptons), which shows that~(i) asymmetry between modes which are charge conjugated each other (CP Asymmetry), and (ii) Lepton Number Violation~(LNV) and/or Lepton Flavour Violation~(LFV), would reveal, respectively, the presence of such phase(s) and that neutrinos are, in fact, Majorana particles. Concerning processes with $\Delta L=2$, rare meson decays (RMD) here studied play an alternative role to the Neutrino-less Double Beta Decay, allowing to extend the neutrino mass range from $\lesssim 100$ MeV to a few GeV. In this line, it is known that CP asymmetry of such processes is maximized when two Quasi-Degenerate Heavy Neutrinos (QDH$\nu$) participate as an on-shell intermediate state~\cite{Cvetic:2012hd,Cvetic:2013eza,Cvetic:2014nla,Dib:2014pga}, producing a resonance as their masses become almost degenerate. The framework within such QDH$\nu$ have been proposed is the Type I Seesaw Mechanism (SS1)~\cite{Minkowski:1977sc,Yanagida:1979as,GellMann:1980vs,Mohapatra:1980yp}, defined by the adition of one SM-fermion singlet (right-handed neutrinos, $\nu_{Ri}$) per generation, resulting in a neutrino mass matrix given, in the basis $\nu=(\nu_L^c,\,\nu_R)$, by
\begin{equation}
\label{ss-mech}
M_\nu=\bpm 0 & Y v \\ (Yv)^T & M_R\epm\,.
\end{equation}
Here, $v=246\GeV$ is the Vacuum Expectation Value (VEV) of the Higgs Field, $Y$ is a $3\times 3$ Yukawa coupling matrix and $M_R$ is a $3\times 3$ mass matrix corresponding to a Majorana mass term; since the neutrino mass sector of SS1 presents a mass matrix (Eq.~\eqref{ss-mech}) contracted with basis which are charge-conjugate each other ($\nu^c$ and $\nu$), we say that the whole mass matrix is a Majorana type one, providing its $M_R$-term the source for explicit LNV. Diagonalization of Eq.~\eqref{ss-mech} provides both light and heavy mass states: the former have masses given by the eigenvalues of $m_\ell^{\textrm{\tiny{SS}}}\sim \nicefrac{(Yv)^2}{M_R}$, whereas the latter have masses $M_h\sim M_R$. The only restriction over $Y$ and $M_R$ is that they have to reproduce the magnitude of light neutrinos masses, $(Yv)^2/M_R\sim 0.1\eV$. In particular, there is a minimal extension to the SM, called $\nu$MSM~\cite{Asaka:2005an,Asaka:2005pn,Gorbunov:2007ak,Shaposhnikov:2006nn}), whose main features are~(i)~masses of SM neutrinos are due to a small Yukawa coupling $Y\sim 10^{-8}$ and $M_R\gtrsim 10^2\MeV$, (ii)~one of the heavy neutrinos, whose mass is in the $\mO(10)\KeV$ range, becomes a candidate for Dark Matter~(DM), and (iii)~the masses of the other two heavy states, lying in the range $M_h\gtrsim 100\MeV$, are close enough to produce the above mentioned effect for RMD. Recently has been proposed by the CERN-SPS Collaboration~\cite{Bonivento:2013jag} to search for Heavy Neutral Leptons (HNL) from rare decay of $B$, $B_c$, $K$ and, preferentially, $D$ and $D_s$ mesons.\\

From an experimental perspective, the drawbacks of seesaw mechanism (type I, as well as types II~\cite{Gelmini:1980re,Magg:1980ut,Mohapatra:1980yp,Cheng:1980qt,Chen:2010uc} and III~\cite{Foot:1988aq}) is that they require values of $M$ to be very large in order to reproduce $m_\ell\sim\mathcal{O}(0.1)\eV$. In fact, assuming that $Y$ lie in the range of Yukawa coupling for SM fermions (i.e., $Y\sim 10^{-6}-10^{0}$) we obtain that $M\sim 10^3-10^{15}\GeV$, and, consequently, that the mixing between SM flavour states and heavy neutrino mass ones is $\tfrac{Yv}{M}\sim 10^{-6}-10^{-11}$, so any manifestation of such heavy neutrinos is out of reach of current reactors. Low Scale Seesaw Mechanism~(LSS) models add not one but two right-handed neutrinos per family to SM ($\nu_{Ri}$ and $S_i$). As only the mass terms $\nu_L\nu_R$ and $\nu_RS$ are allowed, the neutrino mass matrix is given, in the basis $(\nu_L^c,\, \nu_R,\, S)$, by
\begin{equation}
\label{lowss_active-massless}
M_\nu=\bpm 0 & Yv & 0\\ (Yv)^T & 0 & M\\ 0 & M^T & 0\epm
\end{equation}
which results in massless active neutrinos. Including the block $(M_\nu)_{33}=\mu$, active neutrinos acquire masses given by the eigenvalues of $m_\ell^{\textrm{\tiny{ISS}}}\sim (Yv)^2\mu M^{-2}$, which is known as Inverse Seesaw Mechanism (ISS)~\cite{Mohapatra:1986bd}, whereas if we add the block $(M_\nu)_{13}=(M_\nu)_{31}^T=\varepsilon$ they get masses given by $m_\ell^{\textrm{\tiny{LSS}}}\sim (Yv)\varepsilon M^{-1}$, which is called the Linear Seesaw Mechanism (LS)~\cite{Malinsky:2005bi}. We see that required values for $m_\ell$ can be adjusted not only with a big $M$ but with a small $\mu$ or $\varepsilon$, respectively. In fact, using $Y\sim 10^{-1}$, $\mu\sim \mathcal{O}(100)\eV$ or $\varepsilon\sim 10\eV$ we obtain that the heavy parameter of these models ($M$) is not larger than $\sim \mathcal{O}(1)\TeV$, and that the typical mixing with SM particles is $\tfrac{Yv}{M}\sim 10^{-1}$, therefore this models predict phenomenology which could be detected in the short term. In both cases there are two trios of heavy states, $N_1$ and $N_2$, whose masses are the eigenvalues of $M_{1,2}\sim\sqrt{M^2+(Yv)^2}$. \\

The key for obtaining two QDH$\nu$ in $\nu$MSM lies in the fact that when its masses are exactly equal (and, by construction, the otherwise $\KeV$ mass vanishes) the model presents a global U(1) symmetry; so, slightly breaking this symmetry we obtain both the $\KeV$ mass and the quasi-degeneration between the states with $M\gtrsim 100\MeV$~\cite{Shaposhnikov:2006nn}. Furthermore, as this quasi-degeneration comes from the removal of a symmetry in the Lagrangean, its smallness is protected from radiative corrections (we say they are naturally small, in the~'t~Hooft sense~\citep{'tHooft:1979bh}). On the other hand, the mechanism for obtaining an enhancement in the CP asymmetry between a process $P$ and its charge conjugate $P^c$ is, essentially, that their amplitude is the sum of two diagrams, each one of them mediated by Majorana neutrinos with masses $M_1$ and $M_2$. Then, assuming that (i) such QDH$\nu$ are on-shell, and (ii) their interactions with SM particles are very weak (i.e., the Narrow Width Approximation -NWA-), we obtain both the amplitude of Rare Meson Decays (RMD) and its CP asymmetry is maximized when $\Delta m_N=m_{N_2}-m_{N_1}\sim \Gamma_N$, where $\Gamma_{N}$ is the decay width of the heavy neutrino $N$~\cite{Cvetic:2013eza,Cvetic:2014nla} ($\Gamma_N$ is a soft function of $m_N$~\cite{Atre:2009rg}).\\

In this paper we propose that scenarios with Low Scale Seesaw (LSS), i.e., those obtained when we add blocks $(M_\nu)_{33}=\mu$ or $(M_\nu)_{13}=(M_\nu)_{31}^T=\varepsilon$ to Eq.~\eqref{lowss_active-massless}, can provide not one but three pairs of QDH$\nu$, which could enhance the branching ratio (BR) of RMD of mesons going to two charged leptons and another meson, and, eventually, the CP asymmetry between modes which are charge conjugated to each other. For this purpose we need that, in principle, all the intermediate heavy neutrinos are on shell (i.e., $m_M-m_{\ell1}>m_N>m_{M^\prime}+m_{\ell2}$), so its masses have to be in the range of $\mathcal{O}(100)\MeV$ for $K$ decay, and in the range of $\mathcal{O}(1)\GeV$ for $B$ and $D$ decays. Since LSS models propose that masses of heavy neutrinos can be as big as $\sim$TeV scale, we regard scenarios where at least one pair of QDH$\nu$ lie in the range of $\mathcal{O}(10^{-2}-1)\TeV$, so they could contribute to processes testable at LHC.\\

The program of this paper is the following: in Section~\ref{s_proposal} we explain how to get three pairs of QDH$\nu$ in a scenario of Low Scale Seesaw; in Sections~\ref{s_meson-3pair} and~\ref{s-results} we present the formalism for meson decays mediated by three pairs of quasi-degenerate heavy neutrinos and its corresponding results for $B$, $B_c$, $D$ and $D_s$ cases, respectively. Finally, in section~\ref{s_conclusion} we present our conclusions.
\section{Proposal}\label{s_proposal}
\begin{figure}[H] 
\center
\includegraphics[width=35mm, angle=-90]{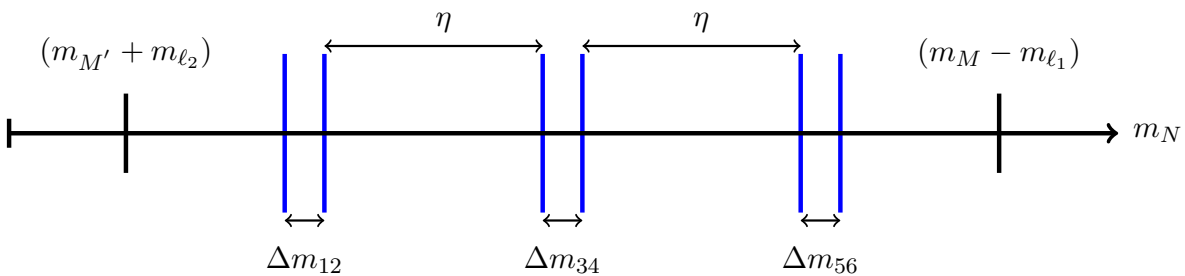}
\caption{Schematic representation of the pairs distribution in on-shell mass range for Scenario III.}  
\label{fig:3pairs}  
\end{figure}

We consider the LSS extension of SM consisting of two families of sterile neutrinos, $\{\nu_{Ri}\}$ and $\{S_i\}$ (with $i=e,\mu,\tau$) \cite{Wyler:1982dd,Witten:1985xc,Mohapatra:1986bd}, which yields, in principle, the blocks of Eq.~\eqref{lowss_active-massless} (up to here, active neutrinos remain massless). Then, generating the term $\tfrac{1}{2}\mu_{ij} \SB{S_i^c}S_j$ ($\mu$-term) or $\varepsilon_{ij}\SB{\nu_{Li}}S_j$ ($\varepsilon$-term) we obtain masses for active neutrinos according to Inverse and Linear seesaw regimes, respectively. With this, we express neutrino mass sector in both flavour basis $(\nu_L^c,\nu_R,S)$ or mass basis $(\nu_\ell,N_1,N_2)$ according to
\begin{equation}
\label{lowss-active-withmass}
\begin{split}
\cL_\txty{mass}^\nu &= m_D\SB{\nu_L}\nu_R+M\SB{\nu_R^c}S+[\mu\textrm{-term or }\varepsilon\textrm{-term}]+ \hc\\
&= m_\ell\SB{\nu_\ell^c}\nu_\ell+m_{1}\SB{N_1^c}N_1+m_2\SB{N_2^c}N_2\\
&=m_\ell\SB{\nu_\ell^c}\nu_\ell+m_1(\SB{N_1^c}N_1+\SB{N_2^c}N_2)+\Delta\SB{N_2^c}N_2\\
\end{split}
\end{equation}
where $m_D\propto vY$ and $\Delta=m_{2}-m_{1}$ (all of them are $3\times 3$ matrices). Now, following the reasoning of Ref.~\cite{Shaposhnikov:2006nn} we can fix $\Delta=0$, obtaining three pairs of exactly degenerate heavy neutrinos (its masses are given by the eigenvalues of $m_{1}\simeq M$; i.e., we have three pairs of { exactly} degenerate heavy neutrinos); after this, we produce a tiny term $\Delta^\prime \SB{N_2^c}N_2$, with $\Delta^\prime\ll m_1$, which causes the spectrum in the $N_{1,2}$ states to become quasi-degenerate. Then, we ask about the symmetry that we have lost in the transition from $\Delta=0$ case to the one in Eq.~\eqref{lowss-active-withmass}. In fact, noting that the second term in the third line of Eq.~\eqref{lowss-active-withmass} can be written as $m_1\SB{\Psi^c}\Psi$, where $\Psi=N_1+N_2^c$ (they have the same absolute eigenvalues), we can stablish that states $\nu_\ell$ and $\Psi=N_1+N_2^c$ have definite charges $(q_\ell,q_\Psi)=(0,\neq 0)$ under certain group U(1), so the operation
\begin{equation}
\label{u1-trans-sobre-n}
\begin{split}
\nu_\ell &\to e^{\imath q_\ell \alpha}\nu_\ell\hs{10}\\
\Psi &\to e^{\imath q_\Psi \alpha}\Psi\,,
\end{split}
\end{equation}
where $\alpha$ is some global phase, leaves invariant $\left.\cL_\txty{mass}^\nu\right|_{\Delta=0}$. Thus, the inclusion of $\Delta^\prime \SB{\Psi^c}\Psi$, with $\Delta^\prime\ll M$, slightly spoils this symmetry (it is clear that $q_{N_2}=-q_\Psi$). Besides, the fact that $\Delta^\prime=0$ restores a (global) symmetry in the Lagrangean is enough reason to expect $\Delta^\prime$ to be naturally small, in the~'t~Hooft sense \citep{'tHooft:1979bh}, i.e., that RGE does not affect its smallness. 
\\\\
Therefore, as we know how to get QDH$\nu$, we shall regard three possible scenarios, depending on how many of them can mediate as on-shell particles in the RMD above mentioned. For this purpose we require that the masses $m_{N_i}$ of all the intermediate neutrinos lie in the range~\cite{Cvetic:2013eza}
\begin{equation}
\label{onshell-region}
m_{M^\prime}+m_{\ell_2} \leq m_{N_i} \leq m_{M}-m_{\ell_1}
\end{equation}
where $\ell_i$ are the final charged leptons. The Scenario~I includes only one on-shell pair and does not offer new phenomenology (in GeV neutrino-mass scale) with respect to the one proposed in $\nu$MSM model \cite{Asaka:2005an,Asaka:2005pn}, regarding that masses of quasi-degenerate neutrinos are in the few GeV range which were studied in Ref.~\cite{Gorbunov:2007ak} (see Refs.~\cite{Cvetic:2012hd}-\cite{Cvetic:2014nla} for predictions about RMD). Even when we are not going to explore Scenario~II, the one with two pairs of QDH$\nu$ in the on-shell range, it is important to mention that, as their masses are sufficiently large, both Scenarios~I and~II offer opportunities for searching sterile neutrinos in collider experiments as LHC~\cite{Kom:2011nc,Dev:2013wba,Das:2014jxa,Helo:2013esa,Milanes:2016rzr}~(particularly, for neutrino masses in the order of TeV); however, has been shown (with some caveats) that sterile neutrinos heavier than LHC mass scale not bring out leptogenesis in case observe LNV at LHC \cite{Deppisch:2013jxa}; see \cite{Dev:2014laa}, section 6, for a extra and helpful discussion. Finally, Scenario~III takes into account all the three pairs in the on-shell mass range~(see Fig.~\ref{fig:3pairs}). By simplicity we shall assume that the mass gap $\eta$ between different pairs satisfies~$\eta / m_N  \sim  \Delta m_{\ell}/m_{\ell} $, or $\eta \sim 0.1\,m_N$, where $m_\ell$ is the mass of light neutrinos~(i.e., heavy neutrinos pairs are so degenerated as light ones). In Section~\ref{s-results} we show results of RMD within the assumptions of each one of these scenarios. As mentioned, future experiments as SHiP~\cite{Bonivento:2013jag} will be a meson factory and could explore intermediate particle masses, from ($\approx 106 \MeV$) until ($\approx 6\GeV$) depending on the initial and final states. 
\section{Meson decays mediated by three pairs of QDH$\nu$}\label{s_meson-3pair}

\begin{figure}[H] 
\center
\includegraphics[width=40mm, angle =-90]{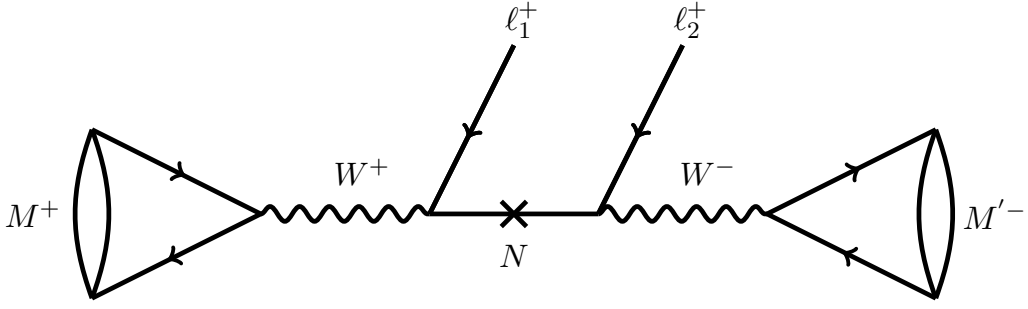}
\caption{The $s$-channel of the lepton number violating decay $M^+ \to \ell_1^+ \ell_2^+ M^{' -}$.}
\label{FigLVM}
\end{figure}

Now we describe the RMD process $M^+ \to \ell_1^+ \ell_2^+M'^{-}$, where $M$ and $M'$ are pseudoscalar mesons: $M=D_s, B_c$ and $M^\prime=\pi, K, D_s$. The most relevant contribution to this decay is shown in Figure~\ref{FigLVM}, and occurs via exchanges of on-shell neutrinos $N_j$. The contributions mediated by off-shell neutrinos and processes including loops ($t$-channel) are strongly suppressed~\cite{Jzamorasaa,Ivanov:2004ch}. Therefore, we focus on the on-shell mass region (Ec.~\eqref{onshell-region}) and tree level processes ($s$-channel).

As we can see in Fig.~\ref{FigLVM}, the process violates the lepton number by two units; in consequence, the intermediate neutrinos ($N_j$) must to be Majorana fermions.

In order to fix notation, we consider that states $\{N_j,N_k\}$ are neutrinos with masses $m_{1},\cdots,m_{N_h}$ ($N_h=6$), where the quasi-degenerate pairs are 12, 34 and 56, whereas the states with arbitrary differences~($\sim 10^{1,2}\MeV$) are 13, 14, 15, 16, 23, 24, 25, 26, 35, 36, 45 and 46. With this, let~\cite{Dib:2014pga}
\begin{equation}
\begin{split}
\mathcal{M}_i&=-G_F^2 V_{q_u q_d} V_{q_u^\prime q_d^\prime} f_M f_{M'} \frac{ B_{\ell_1N_i}B_{\ell_2N_i} m_{N_i}}{p_{N_i}^2-m_{N_i}^2+\imath m_{N_i}\Gamma_{N_i}}\bar{u}(l_2)\slashed{p}_{M^\prime}\slashed{p}_{M}(1-\gamma_5)v(l_1)
\end{split}
\label{amp-decay_with-Ni}
\end{equation}
be the amplitude for the process $M^{+}\to \ell_1^+\ell_2^+ M^{\prime-}$ intermediated by the eigenstate~$N_i$, with~mass~$m_{N_i}$, which enters in the charged current through the mixing $B_{\ell_i N_j}=\sum_\alpha (V^\txty{lep}_{i\alpha })^{*}U_{\alpha j}$  (where $V_{i\alpha}^{\txty{lep}}$ ($U_{\beta j}$) is the matrix element which relate the $i$-th (j-th) charged lepton (neutrino) mass state with $\alpha$-th ($\beta$-th) flavour one). Here we consider that the $B_{\ell N}$ elements includes all the CP-violating phases~\cite{Bilenky:2010zza}. Further, $p_{M}$, $p_{M^\prime}$ are the momenta of mesons $M$, $M^\prime$ and $l_{1}$, $l_{2}$ are the momenta of charged leptons $\ell_{1}$, $\ell_{2}$, whereas $f_{M}$, $f_{M^\prime}$, are the meson decay constant, and $V_{\alpha\beta}$ correspond to CKM element (for instance, if $M$ is a Kaon~$K^+$, $V_{q_u q_d}=V_{us}$).
Thus, the squared amplitude probability for this process is given by
\begin{equation}
\begin{split}
|\mM|^2 &=\sum_{a,b=1}^{N_h}\mM_a^\dagger \mM_b \\
&= \sum_{i=1}^{N_h}|\mM_i|^2+\sum_{\substack{i=1,3,5\\i^\prime =i+1}} \left(\mM_i^\dagger \mM_{i^\prime}+\mM_{i^\prime}^\dagger \mM_{i}\right)+ \sum_{\substack{j,k>j\\\txty{ad pairs}}}\left(\mM_j^\dagger \mM_{k}+\mM_{k}^\dagger\mM_{j}\right)\\
&=\sum_{i=1}^{N_h}|\mM_i|^2+2\sum_{\substack{i=1,3,5\\i^\prime =i+1}}\textrm{Re}\left[\mM_i^\dagger \mM_{i^\prime}\right]+2\sum_{\substack{j,k>j \\ \txty{ad pairs}}}\textrm{Re}\left[\mM_j^\dagger \mM_{k}\right]\,,
\end{split}
\label{total-prob}
\end{equation}
where ad pairs refers to neutrino pairs which have arbitrarily different masses~($13,14,15,\cdots$). Given the fact that heavy neutrinos are weakly interacting particles it is useful to implement the Narrow Width Approximation (NWA),
\begin{equation}
\label{nwa}
\frac{m_{N}}{(p_N^2-m_N^2)^2+(m_N\Gamma_N)^2}\to \pi\frac{\delta(p_N^2-m_N^2)}{\Gamma_N}\,,
\end{equation}
in order to obtain an analytical expression for the terms of Eq.~\eqref{total-prob}. Now, using Eq.~\eqref{nwa} and extending the treatment of decay width for only one pair of QDH$\nu$ we find in Refs.~\cite{Cvetic:2014nla,Cvetic:2013eza}, the corresponding decay width for three pairs is 

\begin{equation}
\begin{split}
\Gamma_\txty{RMD} &=\frac{1}{2!}(2 - \delta_{\ell_1 \ell_2})\frac{1}{2M_M(2\pi)^6}\int d_3|\mM|^2\\
&= 2 (2 - \delta_{\ell_1 \ell_2} ) \Bigg[\sum_{i=1}^{6} |B_{\ell_1 N_i}|^2 |B_{\ell_2 N_i}|^2 \widetilde{\Gamma}_{M}^{(ii)} + 2 |B_{\ell_1 N_1}| |B_{\ell_1 N_2}| |B_{\ell_2 N_1}| |B_{\ell_2 N_2}| \widetilde{\Gamma}_{M}^{(11)} \cos\theta_{21} \delta_{21} \\
& \quad+ 2 |B_{\ell_1 N_3}| |B_{\ell_1 N_4}| |B_{\ell_2 N_3}| |B_{\ell_2 N_4}| \widetilde{\Gamma}_{M}^{(33)} \cos\theta_{43} \delta_{43} + 2 |B_{\ell_1 N_5}| |B_{\ell_1 N_6}| |B_{\ell_2 N_5}| |B_{\ell_2 N_6}| \widetilde{\Gamma}_{M}^{(55)} \cos\theta_{65} \delta_{65} \Bigg]\,.
\end{split}
\label{SplM}
\end{equation}
Here $d_3$ is the number of states available per unit of energy in the final state (3-body phase space) where the factor $(2-\delta_{\ell_1\ell_2})$ refers to the symmetry factor of the amplitude, the factor 2 in front of the latter is due to the contribution of the crossed channel~($\ell_1\leftrightarrow \ell_2$), $\delta_{jk}$ measures the effect of $N_k$-$N_j$ overlap, $\theta_{jk}$ represents the phase difference  $\theta_{jk}  =  (\phi_{1j} + \phi_{2j} - \phi_{1k} - \phi_{2k})$, related with the heavy-light neutrino mixing elements by mean of $B_{\ell_j N_k}  \equiv  |B_{\ell_j N_k}| e^{i \phi_{jk}}$ (where $k,j = 1,2$, see \cite{Cvetic:2014nla}), and, finally

\begin{equation}
\begin{split}
\widetilde{\Gamma}_{M}^{(jj)}&=\frac{K_M^2 m_M^5}{128 \pi^2} \frac{m_{N_j}}{\Gamma_{N_j} } 
\lambda^{1/2}(1, x_j,x_{\ell_1})\times \lambda^{1/2} \left( 1, \frac{x^\prime}{x_j},\frac{x_{\ell_2}}{x_j} \right)\times Q(x_j; x_{\ell_1}, x_{\ell_2},x^\prime) \quad \quad (j=1,...,6)\\
\end{split}
\label{GDDM}
\end{equation}
is the normalized decay width of each sterile neutrino~\cite{Cvetic:2014nla} ($x_j$, $x_{\ell_j}$, functions $\lambda$ and $Q$, coming from integration in $d_3$, are detailed in the Appendix, as well $K_M$). The main difference with the case with only one pair of QDH$\nu$ (Scenario I) is in the three last terms of Eq.~\eqref{SplM}, where we can see the interference within three pairs (instead of one pair) of adjacent heavy neutrinos, $N_1N_2$, $N_3N_4$ and $N_5N_6$. Numerical integrations over the squared amplitude of Eq.~\eqref{total-prob} show that all the other contributions ($13,14,\cdots,46$) are strongly suppressed. As we shall see in \S\ref{s-results}, these contributions will increment the Branching Ratios for RMD, allowing strict restrictions over the couplings $B_{\ell N}$. Besides, in Eq.~\eqref{GDDM} we will assume that $\widetilde{\Gamma}_M^{(jj)}\propto \tfrac{m_N}{\Gamma_N}$ corresponing to adjacent heavy neutrinos (12, 34 and 56) are essentially the same. Therefore, the decay width of pseudo-scalar meson, Eq.~\eqref{SplM}, depends on neutrino masses $m_N$, matrix elements $B_{\ell N}$ and indirectly on the { degeneracy level} $y_{jk}\equiv\frac{\Delta m_{jk}}{\Gamma_{N_j}}$~\cite{Cvetic:2013eza}.
It is important to note that the relation between $\Delta m_{jk}$ and $y_{jk}$ is independent of the already assumed NWA; besides, this $y_{jk}$ enters only indirectly into Eq.~\eqref{GDDM}, through the overlaps $\delta_{jk}$. The latter is manifested implicitly in the parameter $\Gamma_{N_j}$ present in  Eq.~\eqref{GDDM}. Previous studies \cite{Cvetic:2013eza,Cvetic:2015naa,Cvetic:2014nla,Jzamorasaa} have shown that $\delta_{jk}$ is a function of $y_{jk}$ and $y_{jk}=1$ (when $\delta_{jk}=0.5$) is the best choice for measurable CP violation and feasible baryogenesis via leptogenesis \cite{Canetti:2012kh,Canetti:2012vf,Cvetic:2015ura}. From now on we shall assume $y_{jk}=1$ ($\delta_{jk}=0.5$). In addition, we must take into account the acceptance factor, which is defined as the probability of the on-shell neutrino $N_j$ to decay inside the detector of length $L$,
\begin{equation}
P_{N_j} \approx \frac{L}{\gamma_{N_j} \tau_{N_j} \beta_{N_j}} \approx \frac{L\, \Gamma_{N_j}}{\gamma_{N_j}}
\label{PN}
\end{equation}
where $\gamma_{N_j}$ is the Lorentz time dilation factor in the Lab System~{($\sim 2$)}. Consequently, the Effective Branching Ratio~(EBR) is
\begin{equation}
\begin{split}
\textrm{Br}^\txty{eff}(M)=P_{N_j} \textrm{Br}(M)&=P_{N_j} \frac{\Gamma_\txty{RMD}}{\Gamma(M^{\pm} \to \textrm{all})}\,.
\end{split}
\label{BR-eff}
\end{equation}

\section{Results}
\label{s-results}
\begin{figure}[H] 
\subfigure[]{\includegraphics[scale=0.42]{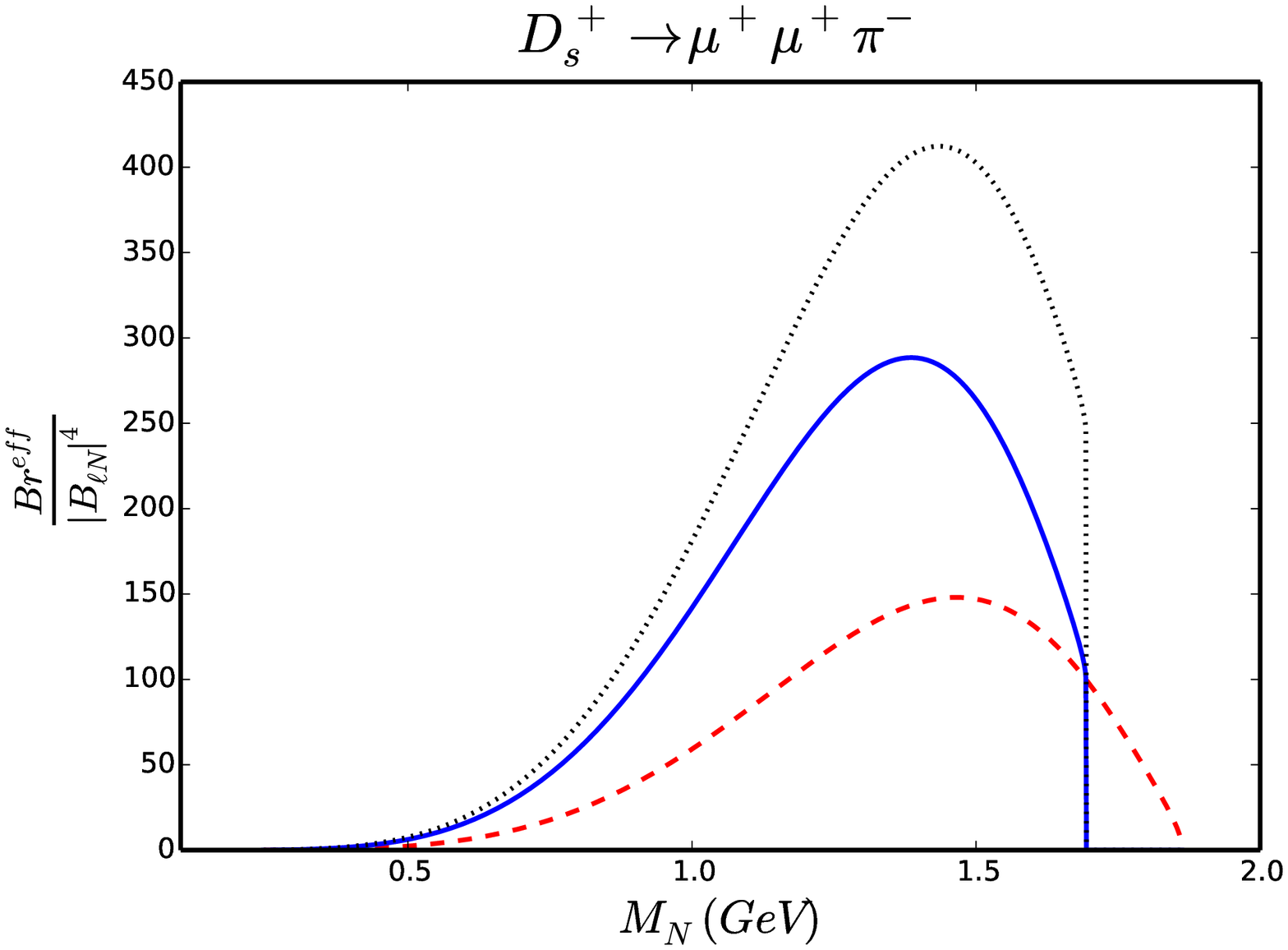}}
\subfigure[]{\includegraphics[scale=0.42]{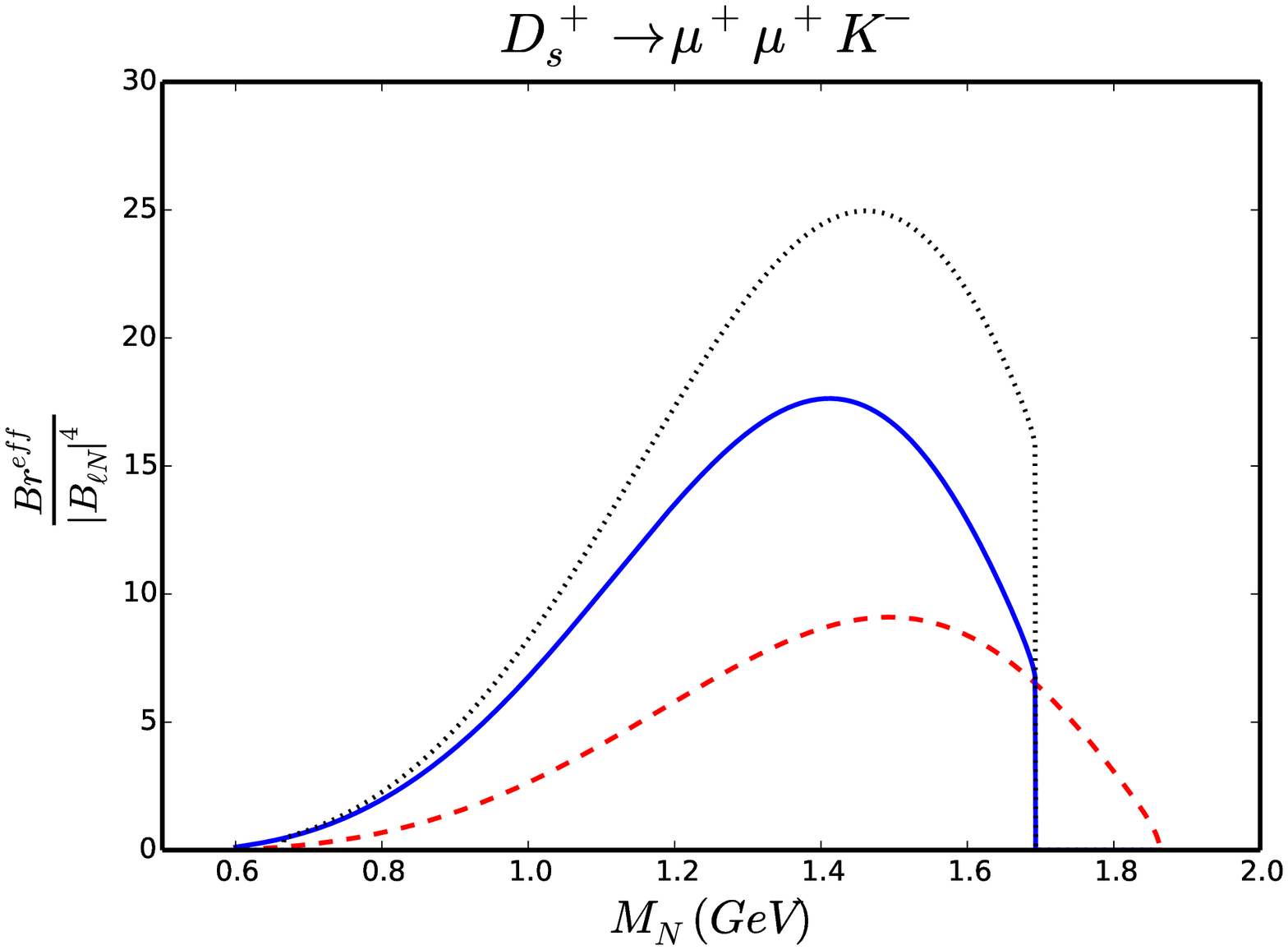}}
\subfigure[]{\includegraphics[scale=0.42]{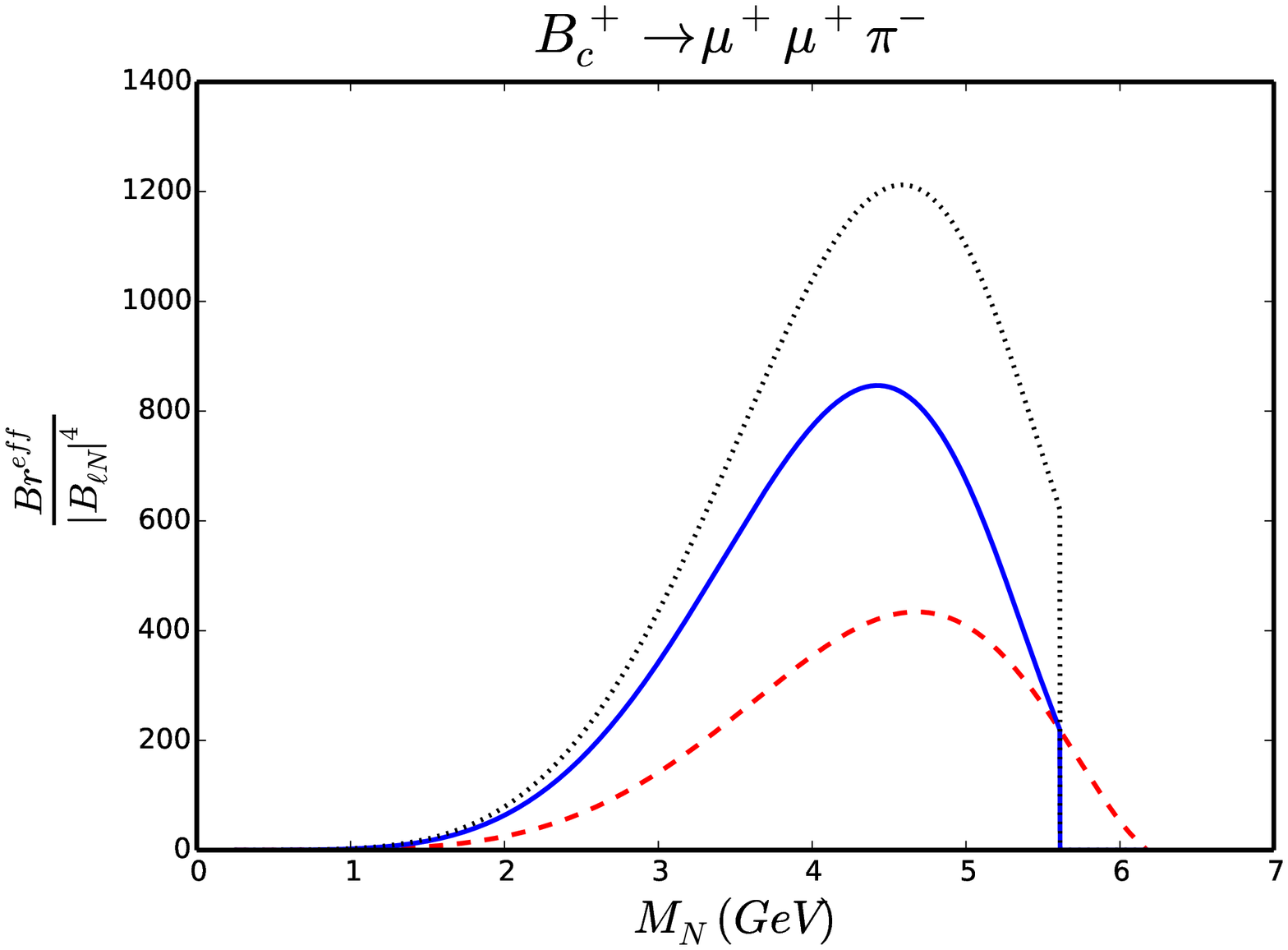}}
\subfigure[]{\includegraphics[scale=0.42]{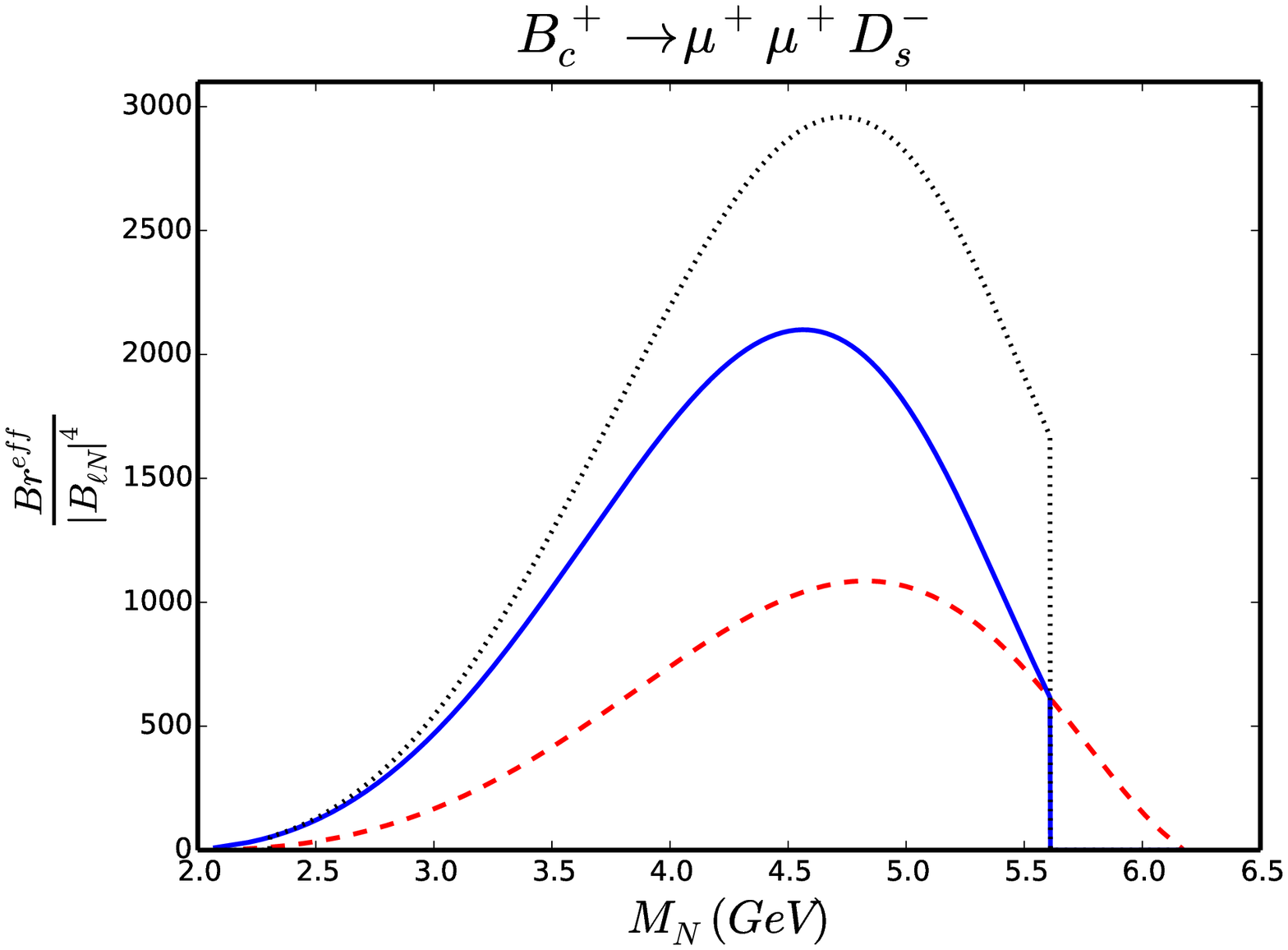}}
\caption{Effective Branching ratio divided by $|B_{\ell N}|^4$, for processes (a)~$D_s^+ \to \mu^+ \mu^+ \pi^-$, (b)~$D_s^+ \to \mu^+ \mu^+ K^-$, (c)~$D_s^+ \to \mu^+ \mu^+ \pi^-$ and (d)~$D_s^+ \to \mu^+ \mu^+ \pi^-$, as function of sterile neutrino mass, for $L=1\m$, $\gamma_N=2$ and $\eta=0.1m_N$. The dashed line (online red) represents the values for scenario I, the solid line (online blue) represents the value for scenario II and the dotted line (online black) represents the values for scenario III. We regard the cases with CP~violation ($\delta_{ij}=0.5$) and $\cos \theta_{jk}=\frac{1}{\sqrt2}$.}
\label{fig:Ds-Bc-decays}
\end{figure}

\begin{figure}[H] 
\subfigure[]{\includegraphics[scale=0.42]{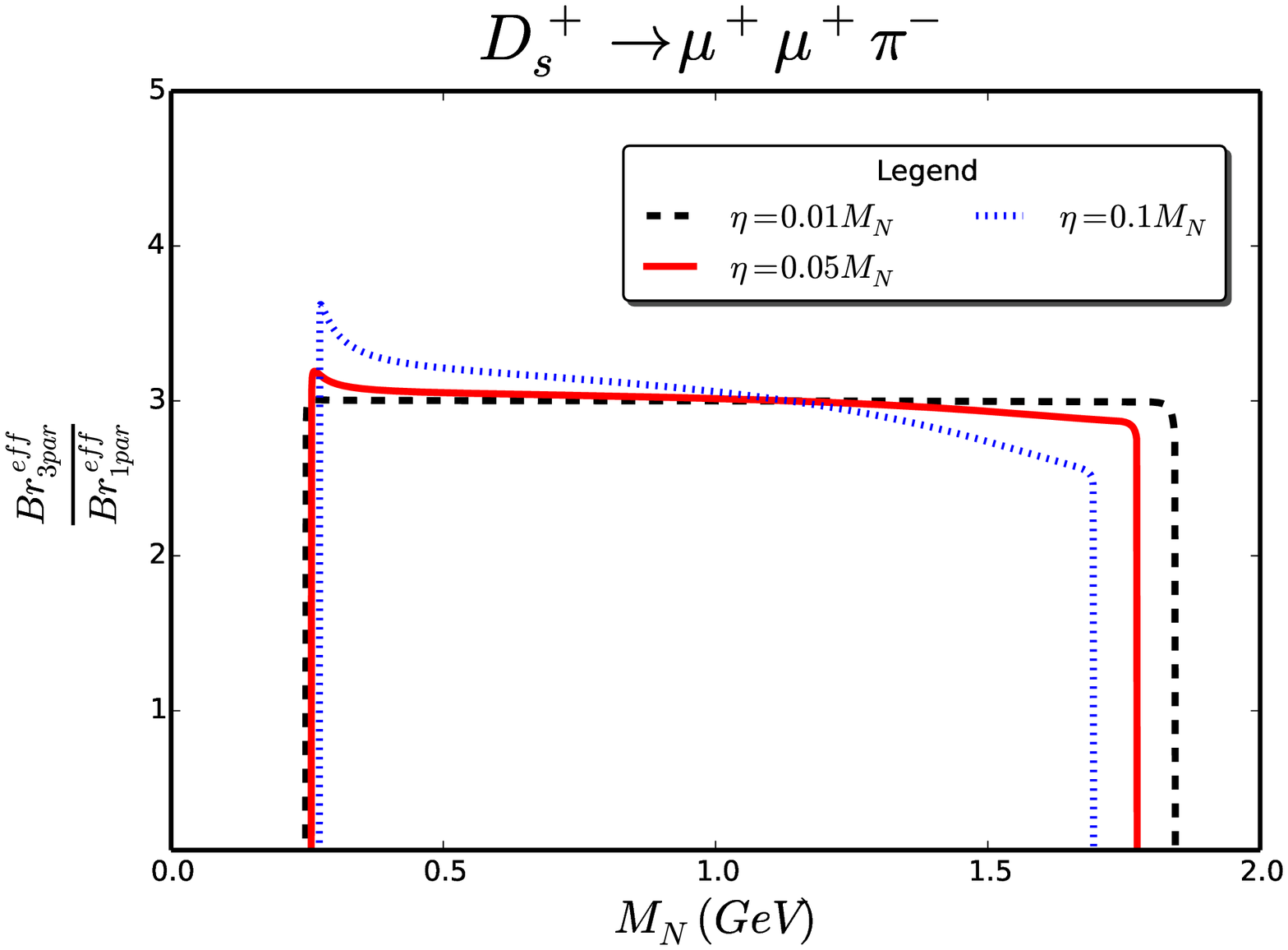}}
\subfigure[]{\includegraphics[scale=0.42]{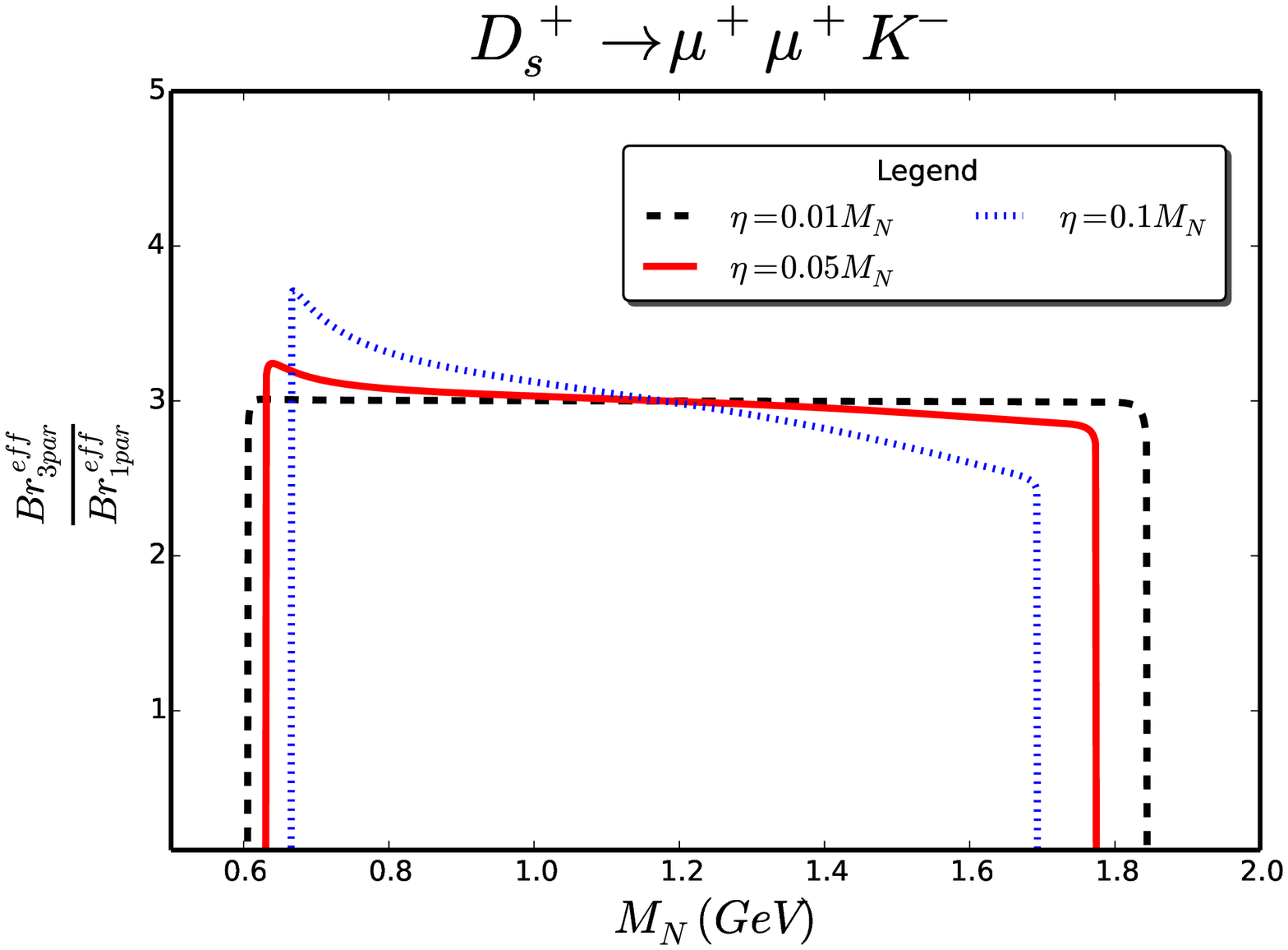}}
\subfigure[]{\includegraphics[scale=0.42]{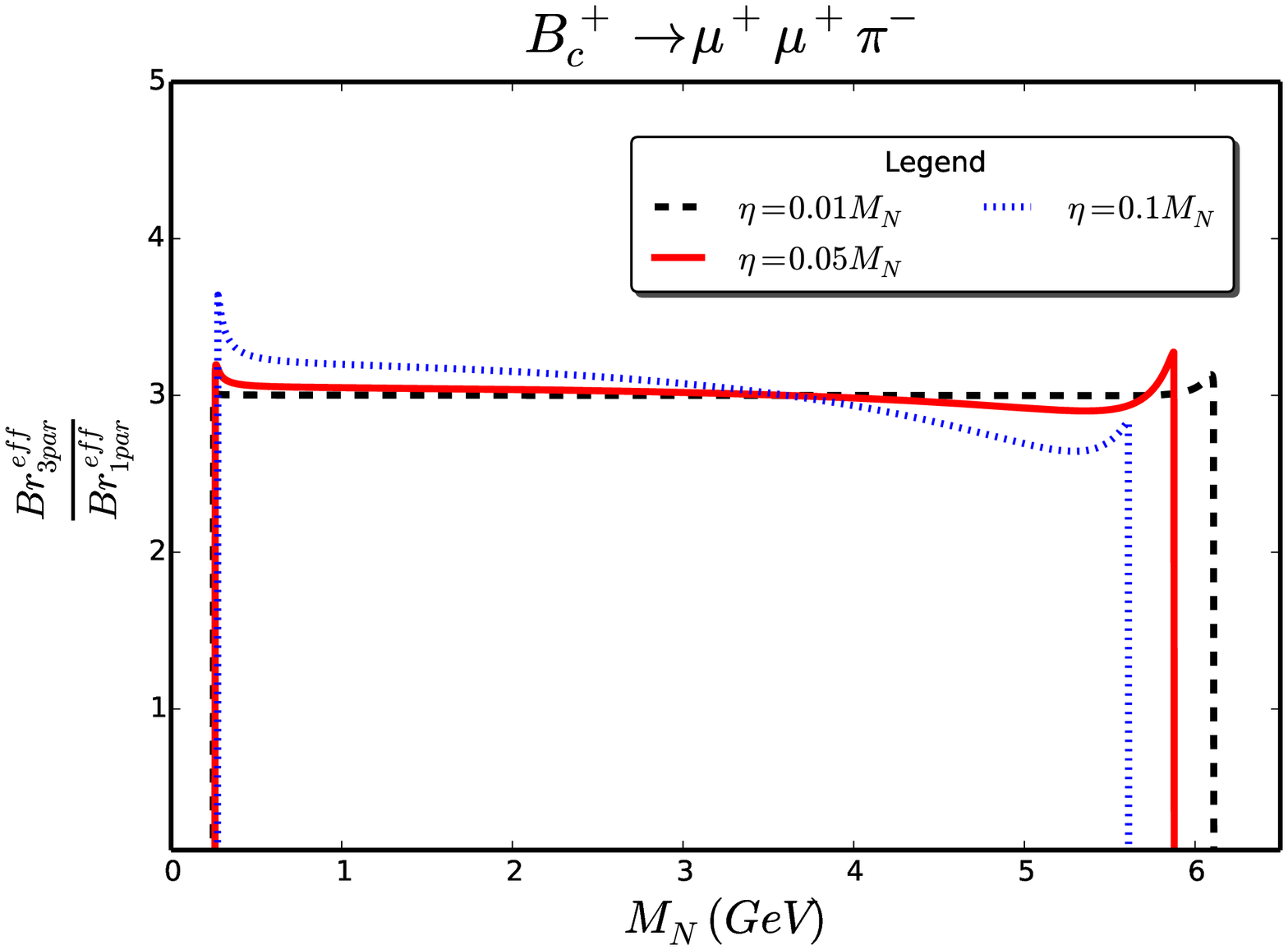}}
\subfigure[]{\includegraphics[scale=0.42]{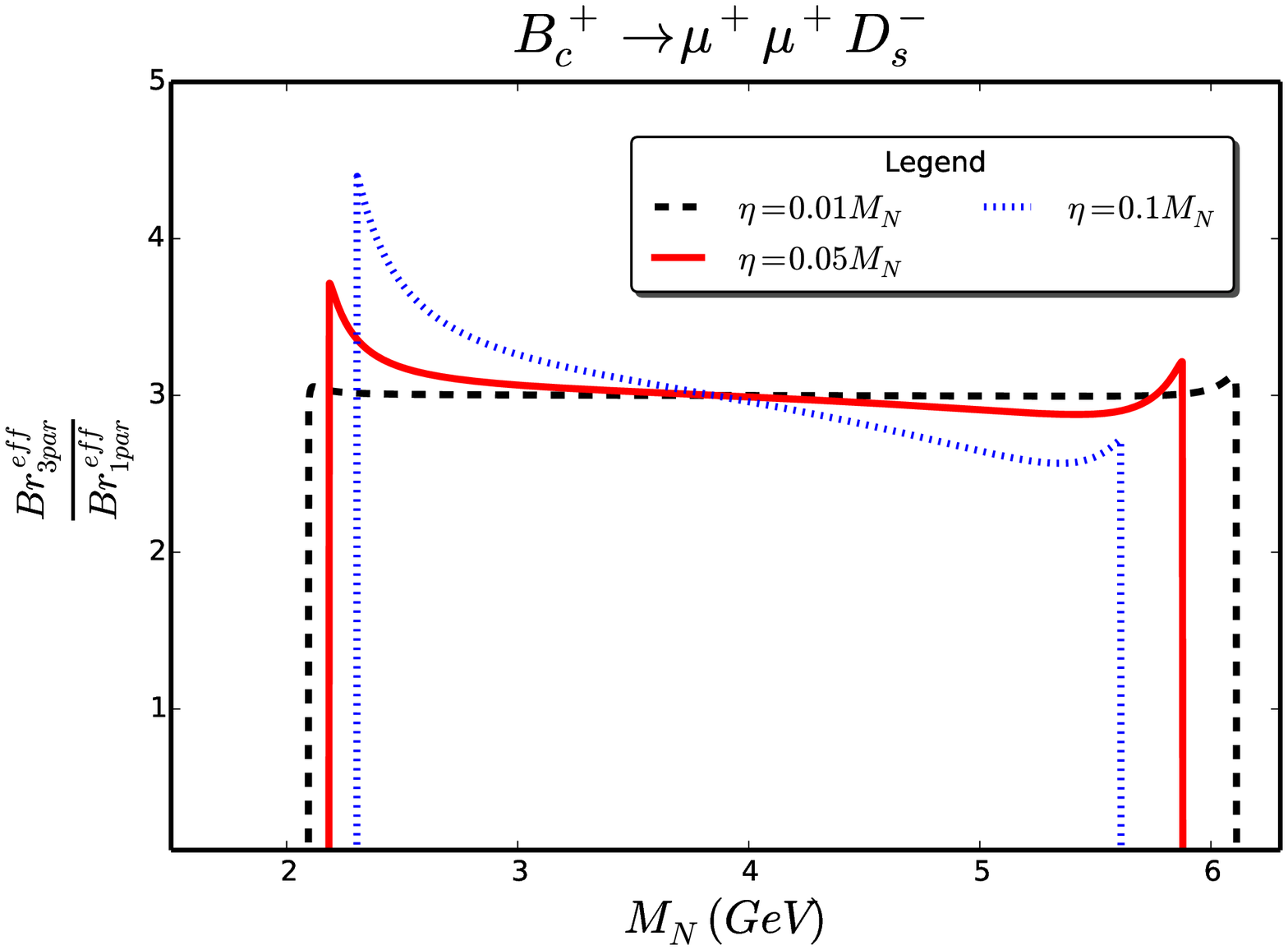}}
\caption{Quotients of the Effective Branching Ratios for different values of $\eta$ parameter using three pairs to one pairs of QDH$\nu$, for processes (a)~$D_s^+ \to \mu^+ \mu^+ \pi^-$ and (b)~$D_s^+ \to \mu^+ \mu^+ K^-$, (c)~$B_c^+ \to \mu^+ \mu^+ \pi^-$ and (d)~$B_c^+ \to \mu^+ \mu^+ D_{s}^-$, as function of sterile neutrino mass. We used $L=1\m$ and $\gamma_N=2$.}
\label{fig:Ds-Bc-coc}
\end{figure}

Now we apply what we know about the decay of mesons mediated by three pairs of on-shell QDH$\nu$ to the processes $D_s^+ \to \mu^+ \mu^+ \pi^-$, $D_s^+ \to \mu^+ \mu^+ K^-$, $B_c^+ \to \mu^+ \mu^+ \pi^-$ and $B_c^+ \to \mu^+ \mu^+ D_s^-$. As we mentioned in previous section, we can deal with three possible scenarios, depending on how many pairs of QDH$\nu$ can mediate as on-shell particles in the~RMD, which depends on whether their masses lie or not in the range of Eq.~\eqref{onshell-region}. By simplicity we shall assume that mass gaps between different pairs $\eta$ satisfy $\eta / m_N  \sim  \Delta m_{\ell}/m_{\ell}$, or $\eta \sim 0.1\ m_N$, where $m_\ell$ represents the masses of active neutrinos. Then, the masses of QDH$\nu$ are labelled as $m_{N_1}=m_N-\eta$, $m_{N_3}=m_N$ and $m_{N_5}=m_N+\eta$, where just the second one will be our independent variable for phenomenological purposes. In consequence, the masses of the heavy neutrinos $(N_1,N_2;N_3,N_4;N_5,N_6)$ are given respectively by $(m_{N_1},m_{N_1}+\Delta m_{12};m_{N_3},m_{N_3}+\Delta m_{34};m_{N_5},m_{N_5}+\Delta m_{56})$. In Figs.~\ref{fig:Ds-Bc-decays} we show the EBR per unit of coupling $|B_{\ell N}|^4$, Eq.~\eqref{BR-eff}, for different meson decays ($M= D_s,B_c$ and $M^\prime=D_s, \pi, K$) assuming $|B_{\mu N_i}| = |B_{\ell N}| $ for $i=1,...,6$ (all equal), as function of such $m_{N_3}=m_N$, regarding all the three scenarios for different initial and final states. In all of them we see that the inclusion of two or three pairs of QDH$\nu$ (scenarios II or III, respectively) results in an increase of the EBR in comparison with the case with only one such pair~\cite{Cvetic:2013eza,Cvetic:2014nla}. In Fig.~\ref{fig:Ds-Bc-coc} we show the ratio between the EBR calculated with three pairs of QDH$\nu$~(EBR3) and the one with only one pair (EBR1) for the decays of Fig.~\ref{fig:Ds-Bc-decays}. In fact, we see even when $m_N$ lies in the range of Eq.~\eqref{onshell-region}~(i.e., [0.25--1.76]\,GeV and [0.60--1.76]\,GeV for $D_s$, [0.25--6.30]\,GeV and [1.98--6.30]\,GeV for $B_c$), the actual ranges for the plots of scenario III in Figs.~\ref{fig:Ds-Bc-decays}-\ref{fig:Ds-Bc-coc} are the ones for which 
\begin{equation}
\frac{m_{M^\prime}+m_{\ell_2}}{1-f}\leq m_N \leq\frac{m_{M}-m_{\ell_1}}{1+f}\,,
\label{onshell-actual-region}
\end{equation} 
where $f=\sfrac{\eta}{m_N}\sim 0.1$, because we demand that all the three pairs contribute to the EBR and, then, to its respective ratios with EBR1~(otherwise we are in scenarios I or II, which are not the goal of this work). This is the reason why the decays of $D_s$ and $B_c$ exhibit an abrupt cut at $m_N\simeq 1.7\GeV$ and $m_N\simeq 5.7\GeV$, respectively. Besides, in Figs.~\ref{fig:Ds-Bc-coc} we see that predictions for EBR3 are between three and four times greater than EBR1. It is interesting to note that, even when these ratios are almost constant in the allowed range for $m_N$, they have a significant increases~(cups) near the extremes. In order to know why this is happening, in Fig.~\ref{fig:Ds-Bc-coc} we show the corresponding ratios for different values of $\eta$~(the degeneracy among QDH$\nu$), and we note that we get smaller increase as we reduce $\eta$. This is easy to understand in the light of Figs.~\ref{fig:3pairs} and~\ref{fig:Ds-Bc-decays}:~(i)~as $m_N=m_{N_3}\simeq m_{N_4}$, the EBR for values of small (great) $m_N$ always get contributions from one pair with masses around $m\sim m_N+\eta$ ($\sim m_N-\eta$), so (ii) only when $\eta$ is sufficiently small all the pairs lie in the extreme zone, given a total EBR corresponding only to extreme masses; (iii) otherwise, when $\eta$ is large an EBR labelled with an extreme mass contains contributions from masses closer to the middle region of Eq.~\eqref{onshell-actual-region}, which clearly yields greater values of EBR. It is worth to mention that when we ignore the $\eta$-effect (i.e., making $\eta\ll m_N$), the EBR3 is just amplified by a factor three with respect to EBR1, hence the shape of dashed lines in the plots of~Fig.~\ref{fig:Ds-Bc-coc}. (This is because all the mass dependence from phase space in Eq.~\eqref{GDDM} is the same for each intermediate heavy neutrino). Finally, we note that $\eta$-effect produces an increase or a decrease of these ratios as $m_N$ is, respectively, smaller or larger than certain $m_N$ (a function of masses of external particles). This can be understood by means of Fig.~\ref{fig:Ds-Bc-decays}, where the peak of EBR3 always occurs for a mass smaller than the mass for which EBR1 has its maximum; therefore, comparing the slopes of EBR1 and EBR3 after its respective maxima, we see that the latter decreases faster than the former, contributing to the decreasing of the ratio in comparison with dashed curves of Fig.~\ref{fig:Ds-Bc-coc}. Also, we note that the interference terms in Eq.~\eqref{SplM} do not seem to manifest in the ratios of Fig.~\ref{fig:Ds-Bc-coc}. This is due to the fact that, as we have one of such interferences in the denominator and three in the numerator, they cancel mutually, surviving only the factor three above mentioned. It is important to point out that our choice of $\theta_{ij}=\pi/4$ in Figs.~\ref{fig:Ds-Bc-decays}-\ref{fig:Ds-Bc-coc} is a necessary condition to maximize CP violation ($y_{jk}=1$, i.e., $\delta_{jk}=0.5$) and simultaneously the decay width presented in Eq.~\eqref{SplM}. It is worth to mention that the exact point of maximal CP violation implies $\delta_{jk}=0.5$ and simultaneously $\cos \theta_{ij}=0$~\cite{Cvetic:2013eza,Cvetic:2014nla}.

\begin{figure}[H] 
\subfigure[]{\includegraphics[scale=0.42]{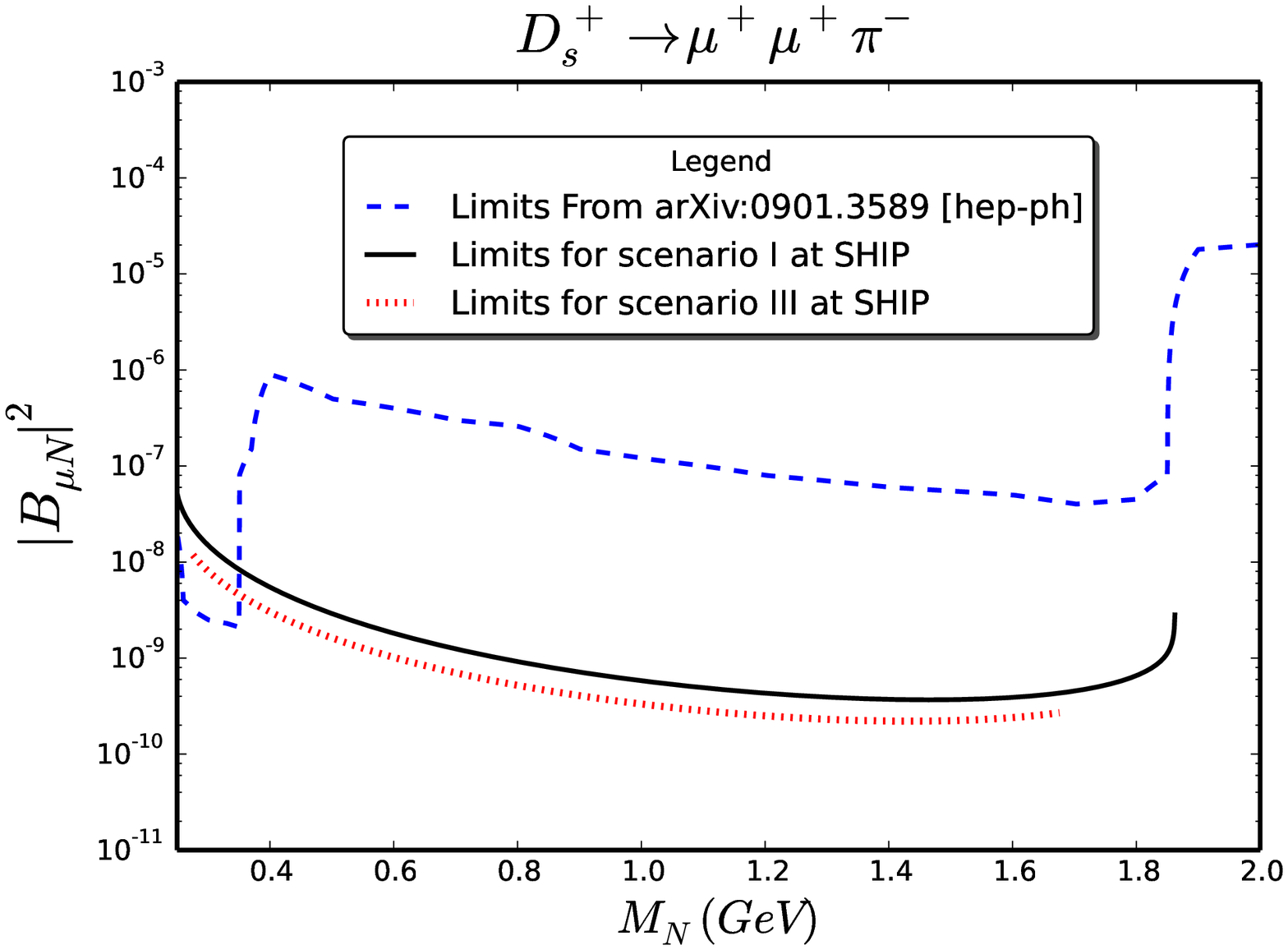}}
\subfigure[]{\includegraphics[scale=0.42]{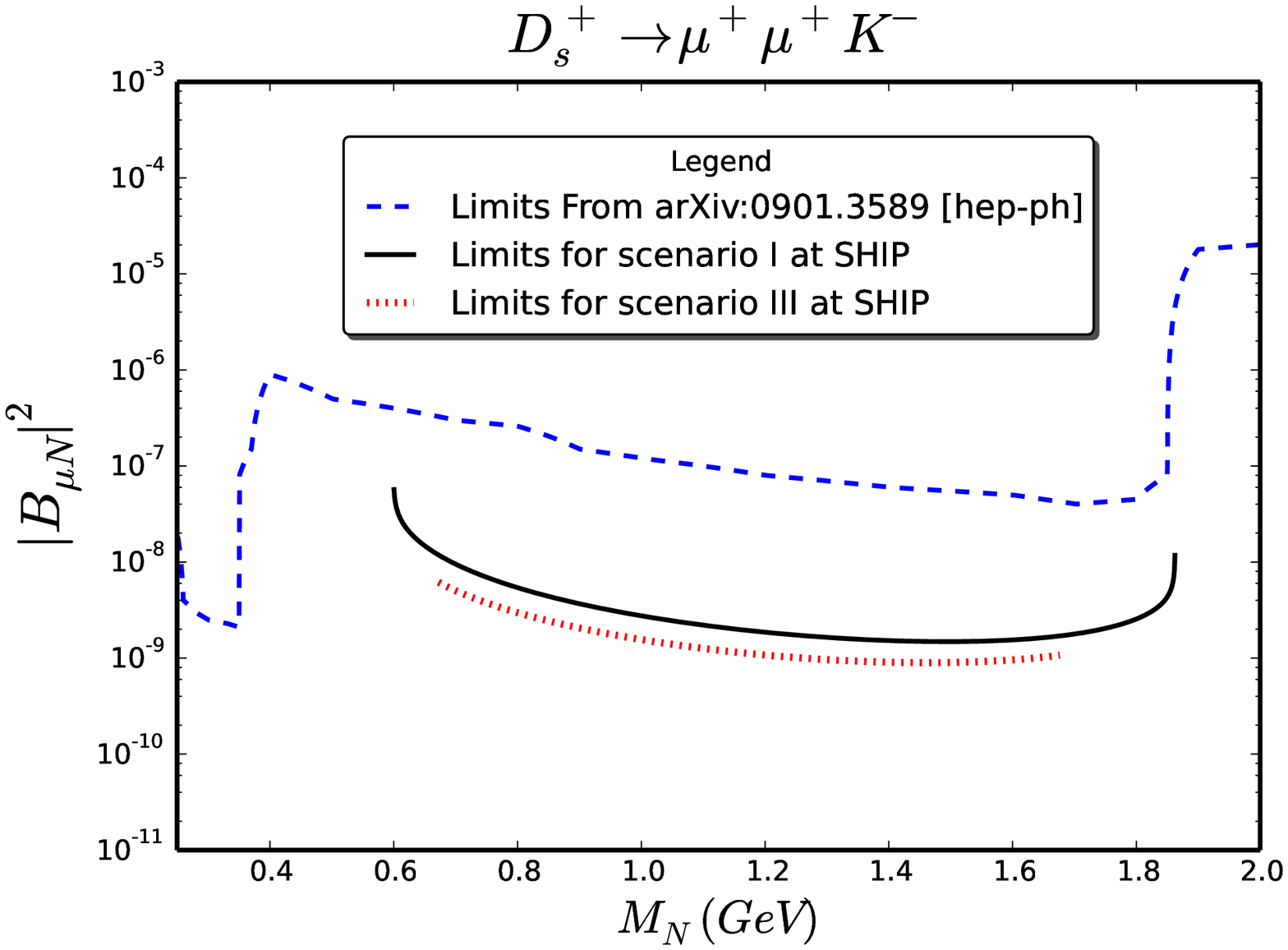}}
\subfigure[]{\includegraphics[scale=0.42]{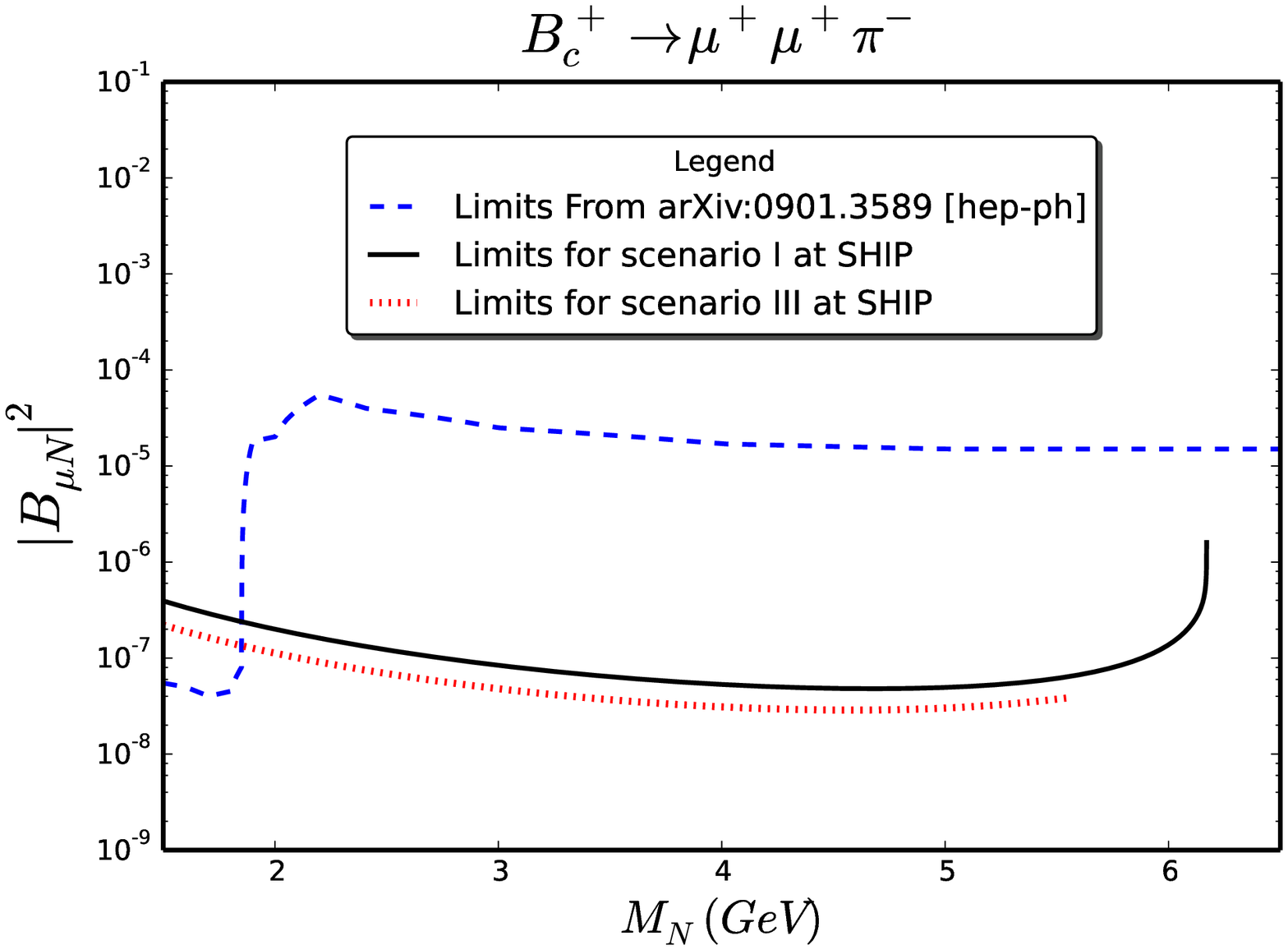}}
\subfigure[]{\includegraphics[scale=0.42]{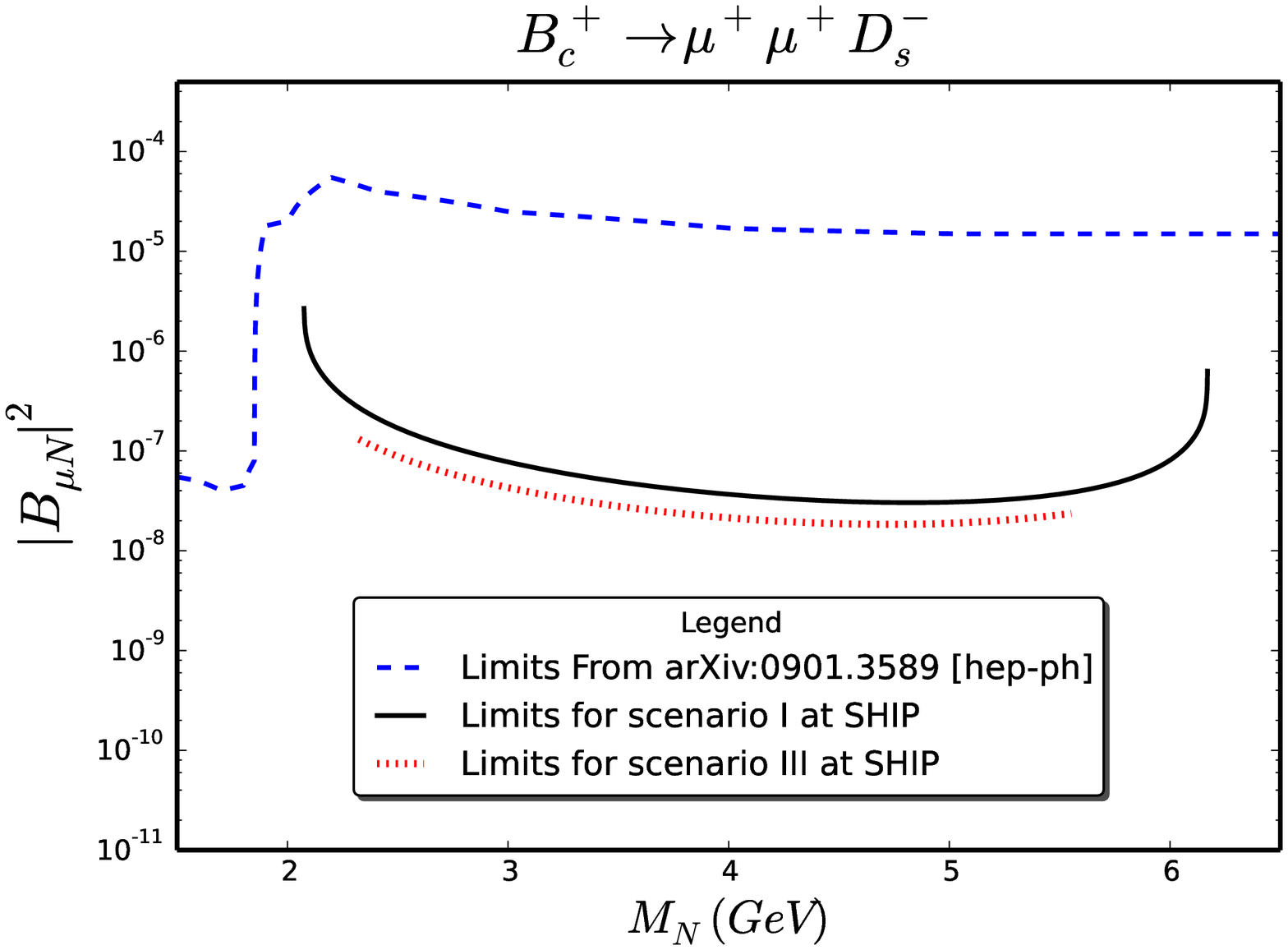}}
\caption{Limits for $|B_{\mu N}|^2$ from Ref.~\cite{Atre:2009rg} vs. the ones we get using one pair and three pairs of QDH$\nu$ for processes~(a)~${D_s^+\to \mu^+\mu^+\pi^-}$, (b)~$D_s^+ \to \mu^+ \mu^+ K^-$, (c)~$B_c^+ \to \mu^+ \mu^+ \pi^-$ and~(d)~$B_c^+ \to \mu^+ \mu^+ D_{s}^-$, based on expected luminosities for mesons. As before, we used $L=1\m$, $\gamma_N=2$ and $\eta=0.1m_N$.}
\label{fig:Ds-Bc-limits}
\end{figure}

Finally, in Fig.~\ref{fig:Ds-Bc-limits} we show a comparison among the current upper limits for $|B_{\mu N}|^2$ provided by Ref.~\cite{Atre:2009rg}~(based on a model with seesaw type~I) and the ones obtainable from the predictions of our Scenarios I and III, again under the assumption that~$|B_{\mu N}|\sim |B_{\ell N}|$; for a extra discussion about heavy-light neutrino mixing see \cite{Deppisch:2015qwa,Antusch:2015mia,Antusch:2016vyf}. This was done by demanding that the number of predicted events for RMD was ${N_\txty{RMD}=(|B_{\mu N}|^4 f_k)\times N_\txty{mes}\geq 1}$, where $N_\txty{mes}$ is the production rate of mesons per year at SHiP~\footnote{M. Drewes (TU Munich) and N. Serra (Zurich U). Private Communication.}, $N_{D_s}\simeq 5.0\times 10^{16}$ and $N_{B_c}\sim 10^{12}$, and $f_k$ is the factor that includes all the kinematics due to each scenario (in fact, $\eta$-effect is present). Therefore, the plots of scenarios 1 and 3 indicate the minimum value of $|B_{\mu N}|^2$ capable of producing one event of RMD. Even when predictions for EBR3 allow smaller limits for $|B_{\mu N}|$, their differences respect to the ones for EBR1 are dominated only by the factor~$\sqrt{3}$ coming from the three pairs of QDH$\nu$, which is even less notable in a logarithmic plot. Now, the fact that these limits are so close implies it will be difficult to decide which underlying seesaw scenario is the origin of these RMD.

\section{Conclusion}
\label{s_conclusion}
We studied the rare decays of mesons ($D_s^+ \to \mu^+ \mu^+ \pi^-$,~$D_s^+ \to \mu^+ \mu^+ K^-$,~$B_c^+ \to \mu^+ \mu^+ \pi^-$ and~$B_c^+ \to \mu^+ \mu^+ D_s^-$) regarding they can produce six on-shell heavy neutrinos with masses in the range of $\sim 1-6\GeV$. For this purpose, we assume the context of a low scale seesaw model constructed with the SM field $\nu_L$ and two extra neutrinos~$\nu_R$ and~$S$, where the mass of light neutrinos ($m_\ell$) is obtained by the introduction of a small parameter in the neutrinos mass matrix ($\mu$ or $\varepsilon$, for inverse and linear seesaw regimes, respectively), allowing that the large scale of the model ($M$), the same as the heavy neutrino masses ($m_N$), lies in the above mentioned range. In order to reproduce the conditions we find in the literature (those leading to maximum CP violation and feasible baryogenesis through leptogenesis) we promoted an argument, based on naturalness, which produces a heavy neutrino mass spectrum with three pairs of quasi-degenerate neutrinos~(Fig.~\ref{fig:3pairs}), where the differences between adjacent masses satisfy $m_{N_2}-m_{N_1}\simeq \Gamma_{N_1}$, $m_{N_4}-m_{N_3}\simeq \Gamma_{N_3}$ and $m_{N_6}-m_{N_5}\simeq \Gamma_{N_5}$, where $\Gamma_{N_i}\sim 10^{-20}\GeV$ are total decay widths of $N_i$'s. In other words, we assumed that heavy neutrinos are particles interacting weakly with SM physics. Likewise, we fixed the difference among pairs of heavy neutrinos, $\eta\sim 0.1 m_N$, such that these pairs have similar relative mass patterns as the active neutrino have. In our calculations we simplified many numerical details concerning the Effective Branching Ratios, making all the couplings between the heavy neutrinos and muon equals, $B_{\mu N_i}=B_{\mu N_j}$. We enhanced CP violation effects by choosing the conditions $y_{jk}=1$, implying that the overlap parameters $\delta_{jk}$ between neutrino resonances become appreciable ($\delta_{jk}=0.5$). For definiteness, we choose the CP-violating phase differences $\phi_i-\phi_j$ such that  $\cos(\phi_i-\phi_j)=\nicefrac{1}{\sqrt2}$ when the overlap between wave functions of heavy neutrinos $N_i$ and $N_j$ is $\delta_{ij}=0.5$. Since the masses of the on-shell heavy neutrinos needed to be in a determined kinematic range related with the masses of the external particles of the decays, we considered three possible scenarios depending on how many pairs this range actually contains, and we obtained a consistent increase in the EBR of RMD as we increment the number of QDH$\nu$. In particular, we conclude that the inclusion of two new pairs of QDH$\nu$ essentially triplicates the EBR of the RMD decay width in comparison with the case with only one pair. Besides, we worked with an effective range for neutrino masses in order to consider all the three pairs, and we found that the ratio between EBR3 and EBR1 was not exactly three, but there was a small variation due to the fact that these pairs were separated by an amount $\eta \leq 0.1 m_N$; this effect vanishes as $\eta\to 0$. The approximate triplication of the EBR we found is consistent with the fact that the mass factor coming from phase space integral is approximately the same, independently of the number of intermediate on-shell neutrino pairs. \\
On the other hand, RMD detection, together with the maximization of CP asymmetry (hence the necessity of QDH$\nu$), is not necessarily attributable only to scenarios like $\nu$MSM, but also to Low Scale Seesaw mechanisms. Furthermore, the latter needs smaller coupling between charged leptons and sterile neutrinos than the former~(Fig.~\ref{fig:Ds-Bc-limits}). Even when our RMDs need neutrinos with masses around a few GeV, it is interesting to note that the off-shell range neutrinos of scenarios~I and~II could perform new phenomenology, regarding that their masses lie in the appropriate range~\cite{Dib:2015oka}. Therefore, there is a phenomenological distinction between this proposal and the one, for instance, of $\nu$MSM: in fact, scenarios~I and~II provide simultaneously QDH$\nu$ pairs of neutrinos which contribute to the EBR of RMD and, besides, heavier neutrinos (now, not necessarily QDH$\nu$) whose phenomenology is testable at~LHC. As a consequence, if experiments find both RMD in a way compatible with QDH$\nu$ and phenomenology of heavier neutrinos in the, say, 100~GeV scale, it could be a signal in favour of LSS mechanism rather than type I seesaw mechanism. Finally, is worth to mention that, for instance, our scenario II provides a couple of quasi-degenerate neutrinos whose masses are still free to be set in the appropriate range in order to contribute to neutrinoless double beta decay, in this context some work have been done \cite{Asaka:2016zib,Drewes:2016lqo,Lisi:2015yma,LopezPavon:2012zg}.
\section{Acknowledgments}
This work was supported by Fellowship Grant {\it Becas Chile} No.~74160012, CONICYT~(J.Z.S), and for Postdoctoral Fellowship Grant No.~3160657, FONDECYT (G.M.). Also, the authors want to thank for valuable discussions with Claudio Dib, Diego Aristiz\'abal and Gorazd Cvetic, as well as the Fellowship Grant {\it Beca Puente} received from CCTVal.

\section{Appendix ``\,Kinematics Functions\,''}
\label{appKF}
The Kinematic Functions shown in Eq.(\ref{GDDM}), coming from the phase space integration, are given by the expressions
\begin{equation}
\begin{split}
\lambda(y_1,y_2,y_3) &=  y_1^2 + y_2^2 + y_3^2 - 2 y_1 y_2 - 2 y_2 y_3 - 2 y_3 y_1\,, \\
Q(x; x_{\ell_1}, x_{\ell_2}, x^\prime) &= {\bigg\{}\frac{1}{2}(x-x_{\ell_1})(x-x_{\ell_2})(1-x-x_{\ell_1})
\left(1-\frac{x^\prime}{x}+\frac{x_{\ell_2}}{x}\right)\\
&\quad+{\big [}-x_{\ell_1} x_{\ell_2}(1+x^\prime +2x-x_{\ell_1}-x_{\ell_2})-x_{\ell_1}^2(x-x^\prime)+x_{\ell_2}^2(1-x) \\
&\quad+x_{\ell_1}(1+x)(x-x^\prime)-x_{\ell_2}(1-x)(x+x^\prime){\big ]}{\bigg \}} \\
&=\frac{1}{2}\left[(1-x)x+x_{\ell_1}(1+2x-x_{\ell_1})\right]\left[x-x^\prime-2x_{\ell_2}-\frac{x_{\ell_2}}{x}  (x^\prime-x_{\ell_2})\right ]
\end{split}
\label{lambdaM}
\end{equation}
where $$x_j= \frac{M_{N_j}^2}{M_M^2}\,, \quad x_{\ell_s} =  \frac{M_{\ell_s}^2}{M_M^2} \,, \quad x^\prime =\frac{M_{M^\prime}^2}{M_M^2} \,,  \quad (j=1,2; \; \ell_s=\ell_1, \ell_2) \,.$$
Since the valence quark content of $M^+$ and $M^{\prime\,-}$ is $q_u {\bar q}_d$ and $q_u^\prime \bar{q}^\prime_d$, respectively, the constants involved in the normalized decay widths of Eq.~\eqref{GDDM} are $$K_{M}=-G_F^2 V_{q_u q_d} V_{q_u^\prime q_d^\prime} f_M f_{M^\prime} \hs{5}\textrm{with}\hs{5} \qquad K_M = (K_M)^{*}\,,$$
where $f_{M}$ and $f_{M^\prime}$ are the meson decay constants of $M^{+}$ and $M^{^\prime-}$,
whereas $V_{q_u q_d}$ and $V_{q_u^\prime q_d^\prime}$ are its CKM elements.

\bibliographystyle{apsrev4-1}

\bibliography{biblio.bib}

\end{document}